\documentclass[twocolumn,prl,amsmath,amssymb,showpacs,superscriptaddress,floatfix]{revtex4}
\usepackage{graphicx}
\usepackage{bm}

\usepackage{epsf}
\begin{document}
\title{A novel anomalous region of water}

\author{Yu. D. Fomin, E. N. Tsiok, V. N. Ryzhov, and V. V. Brazhkin}

\affiliation{ Institute for High Pressure Physics RAS, 108840
Kaluzhskoe shosse, 14, Troitsk, Moscow, Russia}

\date{\today}

\begin{abstract}
Water is the most important liquid in the Universe. At the same
time it is the most anomalous liquid. It demonstrates several
dozens of anomalies, among which are density anomaly, diffusion
anomaly etc. Anomalous behavior of water is a topic numerous
publications. However, most of the publications investigate the
anomalous behavior of water in the vicinity of critical points:
the liquid-gas critical point and the second hypothetical critical
point in supercooled region. Here we analyze experimental data on
such properties of water as heat capacity, speed of sound, dynamic
viscosity and thermal conductivity. We show that these properties
demonstrate anomalous maxima and minima in a region which is far
from both critical points. Therefore, we find a novel region of
anomalous properties of water (anomalous triangle) which cannot be
related to critical fluctuations. We also perform a molecular
dynamics simulations of this region with two common water models -
SPC/E and TIP4P - and show that these models fail to describe the
novel anomalous region.
\end{abstract}


\maketitle

Water is the most important liquid in the Universe. The existence
of life is attributed to water. At the same time water is very
strange liquid which demonstrate numerous anomalous properties
(see, for instance, \cite{water-anom} for the list of anomalies of
water). Many of these properties are responsible for importance
water for life and technology, e.g. density anomaly or extremely
high solubility of many substances in water. All of this made
water one of the most studied substance. However, up to now many
questions still remain unsolved. Moreover, investigations of water
bring some novel discoveries. For example, many novel solid phases
of ice were found in the last few decades. That is why in spite of
numerous efforts water is still a unique universe of problems for
scientists.

As it was mentioned above, water exhibits many anomalous
properties. More then 70 anomalies are described in the
literature. Many of them are well described in a recent review
paper \cite{gallo-tale}. Anomalous properties of water include
such phenomena as density anomaly (negative thermal expansion
coefficient), diffusion anomaly (diffusion coefficient increases
under isothermal compression), structural anomaly (water becomes
more structured under isothermal compression) etc. More anomalies
are related to the behavior of thermodynamic response functions -
heat capacity, isothermal compressibility, isobaric expansion
coefficient etc. These anomalies are often related to the so
called Widom line, which is defined as a line of maxima of
correlation length in the vicinity of critical point
\cite{widom-init}. Many other quantities, e.g. the response
functions mentioned above also demonstrate maxima close to the
critical point and it was supposed that the locations of maxima of
different quantities should be close to each other. However, in
our recent publications it was shown that even in simple liquids
like Lennard-Jones fluid or square-well system the lines of maxima
of different functions rapidly diverge and one cannot approximate
them by a single line \cite{widom-lj,widom-sq}. Moreover, even the
maxima of the same quantity taken along different thermodynamic
paths (for example, isotherms or isobars) can be different
\cite{widom-co2}. Therefore, the Widom line is not uniquely
defined. The Widom line of water in the vicinity of liquid-gas
critical point was reported in Ref. \cite{widom-water}. Another
reference \cite{widom-water-dynam} demonstrated that bends of
transport coefficients of water take place at the line of maxima
of isobaric heat capacity $c_P$ of water.

In Ref. \cite{mish-st} it was proposed that some anomalous
properties of water can be induced by a liquid-liquid critical
point which is located in so-called 'no man's land' - an
experimentally inaccessible region of the phase diagram. The
fluctuations above this critical point induce the maxima of
isobaric heat capacity, thermal compressibility, etc. This concept
attracted a lot of attention of researchers and many experimental
and theoretical works discussed it. In particular, many
publications investigated the second Widom line of supercooled
water (see, for instance, \cite{ws1,ws2,ws3} and references
therein).

One can see that investigation of properties of water is a very
hot topic and the literature on it is extremely vast. However, one
can notice that most of these studies are restricted to relatively
small pressures. Moreover, many theoretical models explaining the
origin of these models are based on the critical fluctuations.
From studies of the Widom line
\cite{widom-lj,widom-sq,widom-co2,widom-water,widom-water-dynam}
one can see that the critical fluctuations extend to relatively
narrow region with $T\leq 2T_c$ and $P\leq 7P_c$. In case of water
the regions of critical fluctuations related to two critical
points can be estimated as $500  <T < 1000$ (temperature in
Kelvins), $200 <P < 1500$ (pressure in bars) for liquid-gas
critical point and $200 < T < 400$ K and $0 < P < 1000$ bar for
liquid-liquid critical point. These regions are shown in Fig. 1.

\begin{figure} \label{regions}
\includegraphics[width=8cm, height=8cm]{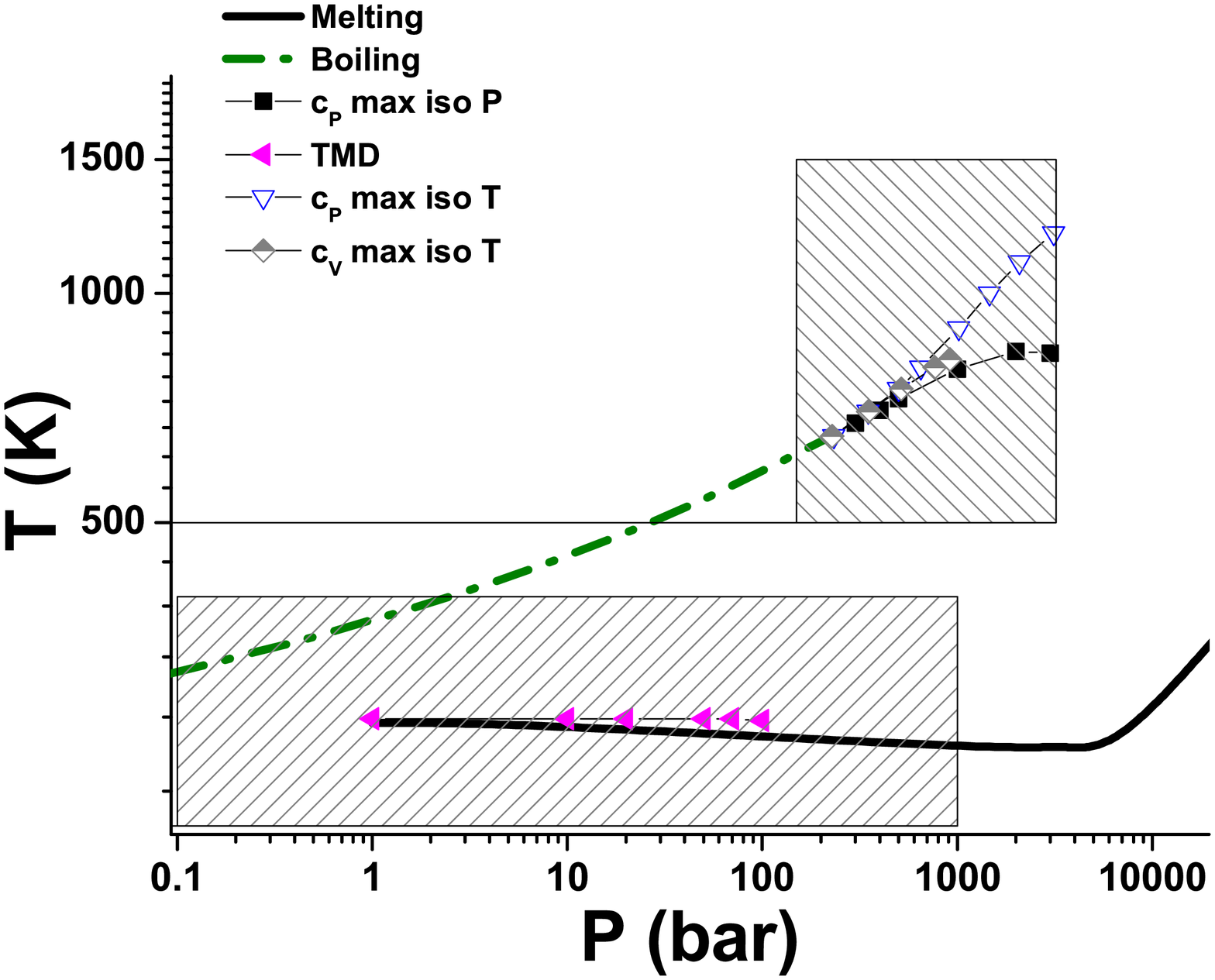}%

\caption{The regions of phase diagram where critical fluctuations
can induce anomalous behavior. TMD means the temperature of
maximum density which represents the density anomaly of water.}
\end{figure}

From Fig. 1 one can see that the investigation of anomalous
behavior of water is restricted to two relatively narrow regions.
In particular, there is no investigations of water anomalies under
high pressure. Nowadays the methods to obtain high pressures and
perform experiments under such conditions rapidly develop and
novel records of pressure are achieved. The behavior of many
substances under pressure appears to be very unexpected which
makes the high pressure studies extremely intriguing and
interesting topic.

The melting line of water is measured up to the pressure of 60
GPa. In the present study we combine the melting line of water
from many publications
\cite{melt1,melt2,melt3,melt4,melt5,melt6,melt7}. However, the
properties of liquid water under such conditions are much less
studied. Even if there are some experimental works on it, often
they are not systematized which make harder to construct a
self-consistent picture of the behavior of water under high
pressure.

In the present paper we discuss the anomalous properties of water
under high pressure. We analyze available experimental results and
compare them to molecular dynamic calculations.

\bigskip

We start our investigation from detailed analysis of experimental
data which are summed up in NIST database \cite{nist}. This
database summarizes numerous experimental data. The database uses
approximations for the experimental data. Several quantities are
reported there. Here we consider such properties of water reported
in NIST database as equation of state (pressure as a function of
density and temperature), isochoric and isobaric heat capacities -
$c_V$ and $c_P$ respectively, adiabatic speed of sound $c_s$,
dynamic viscosity $\eta$ and thermal conductivity $\lambda$. We
consider these quantities in the whole range of thermodynamic
parameters given in the database and find the locations of their
minima and maxima. These points are placed on the phase diagram.
The boiling curve of water is taken in Ref. \cite{boiling-exp} and
the melting curve is combined from the data given in Refs.
\cite{melt1,melt2,melt3,melt4,melt5,melt6,melt7}

In a set of our recent papers we showed that a so called Rosenfeld
relation \cite{10,11} which make a connection between transport
coefficient of a fluid with excess entropy can be valid along
isochors, but break down along isotherms \cite{ros1,ros2}. It led
us to a conclusion that some properties of fluid appear
differently along different trajectories in the space of
thermodynamic variables \cite{traj1,traj2}. Later we discovered
that the locations of maxima of the same quantity in the vicinity
of a critical point also depends on the thermodynamic path
\cite{widom-co2}. That is why it is important to monitor the
quantities of interest along different "trajectories" in the space
of thermodynamic variables. Here we study the all properties along
isochores, isotherm and isobars. The corresponding curves in the
plots are denoted as iso V, iso T and iso P respectively.

Fig. 2 gives some examples of anomalous behavior of several
properties of water under high pressure. The pressure is chosen
$P=4000$ bar. The temperature ranges from $T_{min}=273$ K up to
$T_{max}=1275$ K. One can see that the isobaric heat capacity
$c_P$ (Fig. 2 (a)) demonstrates two maxima at the given pressure:
at $T=318$ K and $T=704$ K. A minimum of $c_P$ is also observed at
$T=420$ K. The isochoric heat capacity $c_V$ (Fig. 2 (b)) has only
a maximum at $T=303$ K which is very close to the low temperature
maximum of $c_P$. The maximum of speed of sound $c_s$ (Fig. 2 (c))
is located at $T=378$ K. The behavior of viscosity at this
pressure is monotonous (not shown) while the thermal conductivity
shown at Fig. 2 (d) has a maximum at $T=526$ K.

\begin{figure} \label{p4000}
\includegraphics[width=6cm]{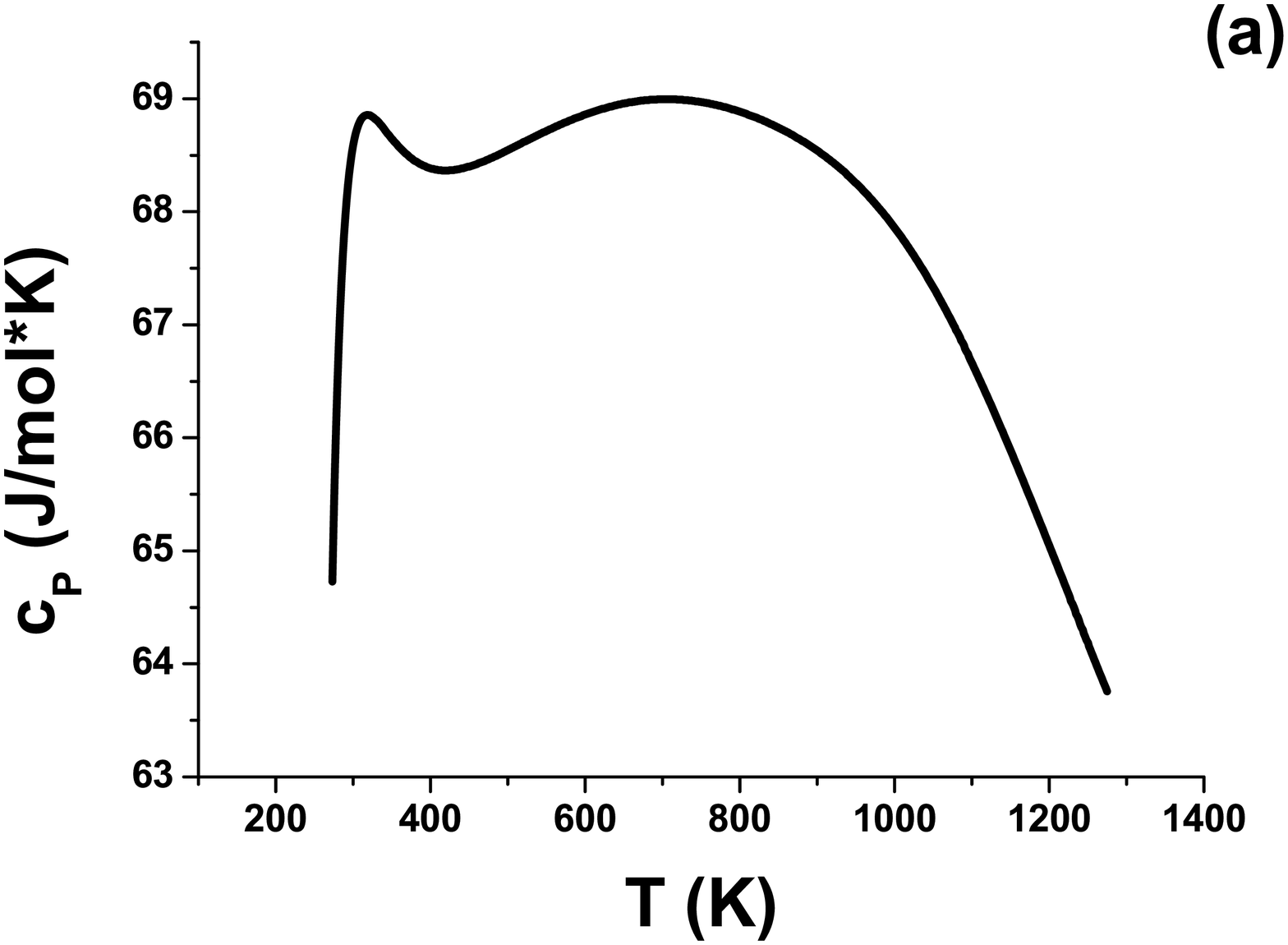}%

\includegraphics[width=6cm]{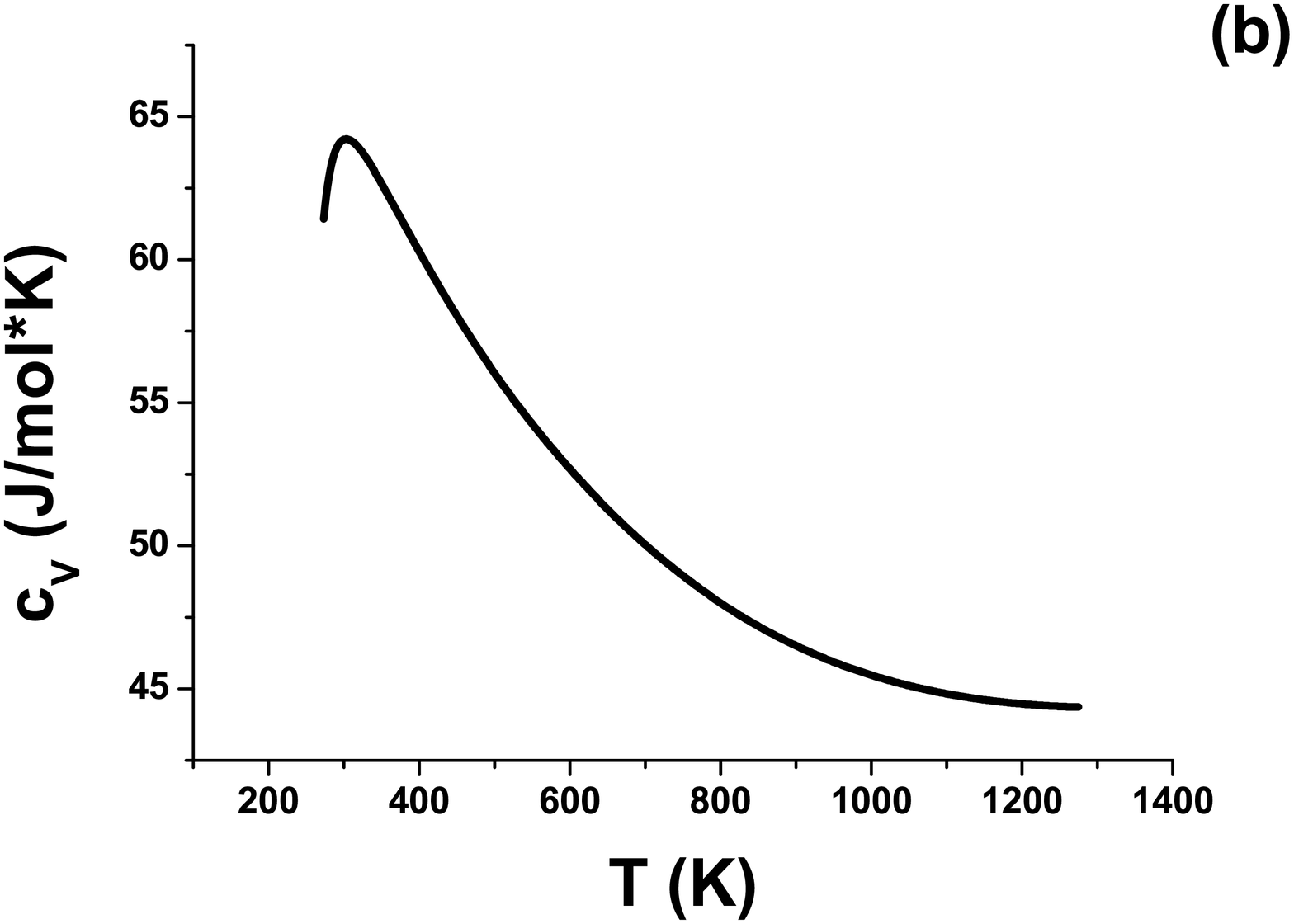}%

\includegraphics[width=6cm]{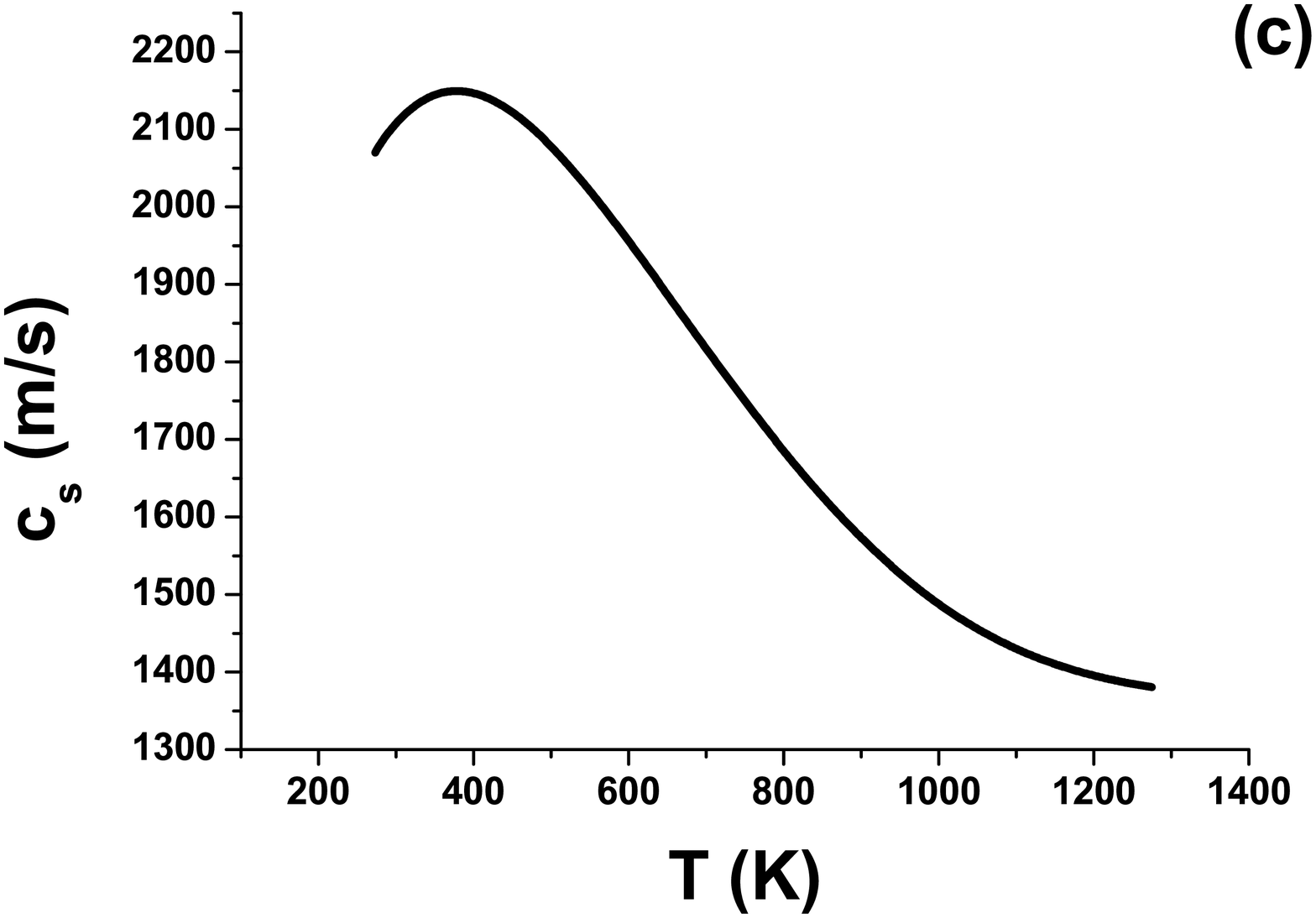}%

\includegraphics[width=6cm]{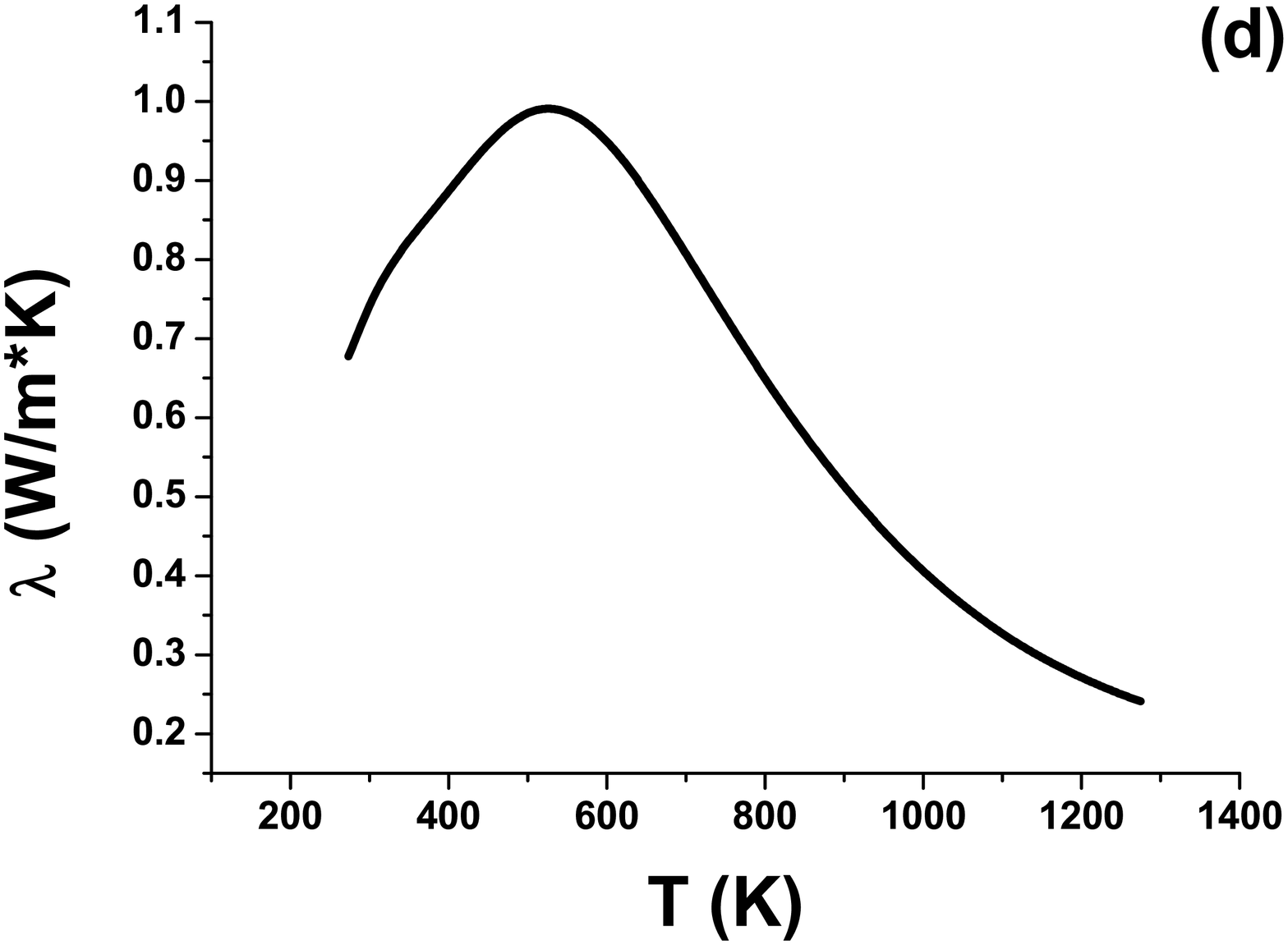}%

\caption{Examples of anomalous behavior of several properties of
water under high pressure. The pressure is $P=4000$ bar (0.4 GPa)
(a) isobaric heat capacity; (b) isochoric heat capacity; (c)
adiabatic speed of sound; (d) thermal conductivity. The data are
taken from NIST database.}
\end{figure}

We have explored the data along many isobars, isotherms and
isochors and located minima and maxima of different quantities.
The results are given in Fig. 3. Fig. 3 (a) gives the overall
picture, while panels (b) and (c) enlarge the region close to the
critical point and to the bend of the melting line respectively.
Note, that some quantities demonstrate several maxima or minima
along the same trajectory as we saw for $c_P$ along $P=4000$ bar
isobar. The melting and boiling lines are also given.

\begin{figure} \label{anom-all}
\includegraphics[width=6cm, height=6cm]{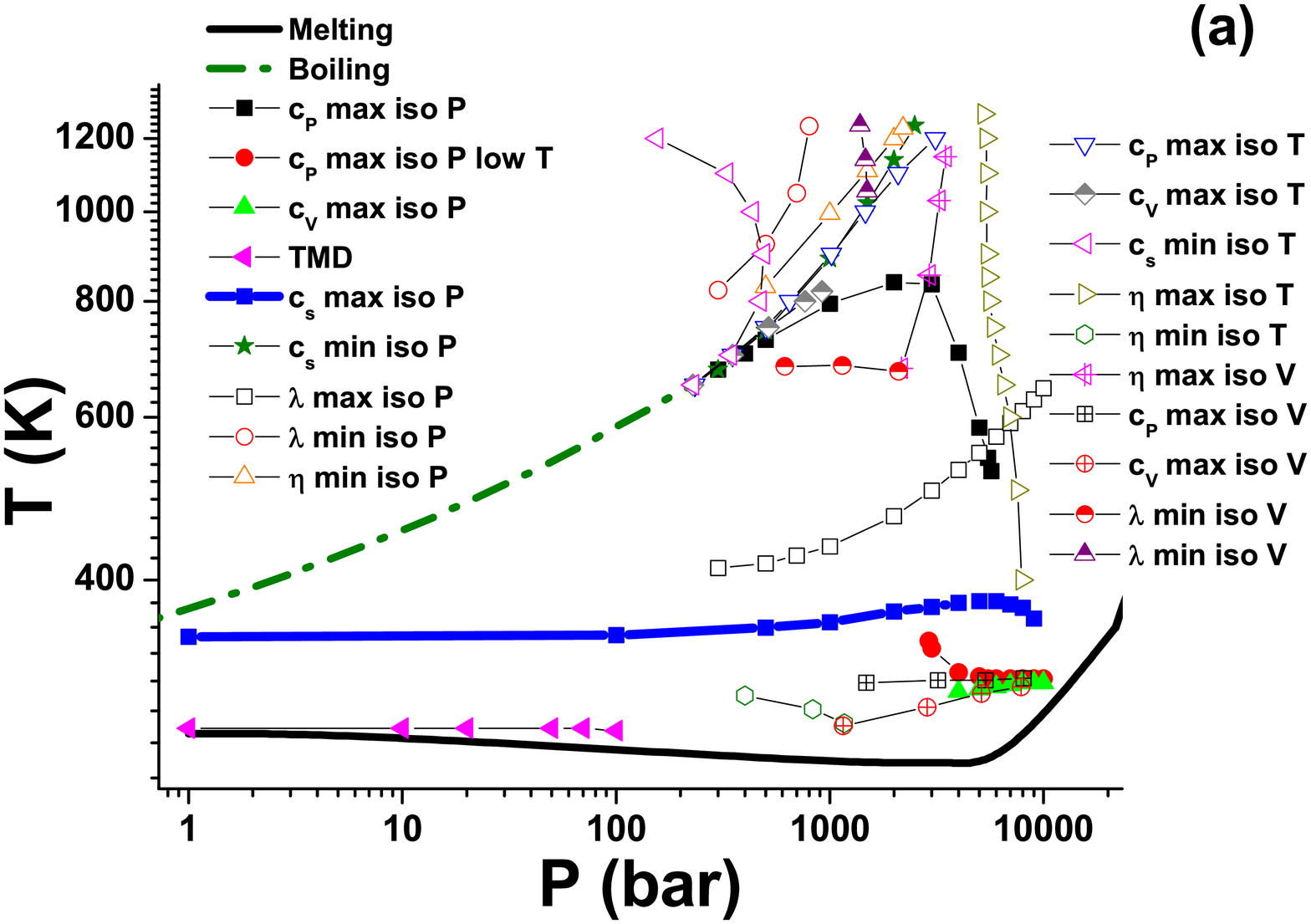}%

\includegraphics[width=6cm, height=6cm]{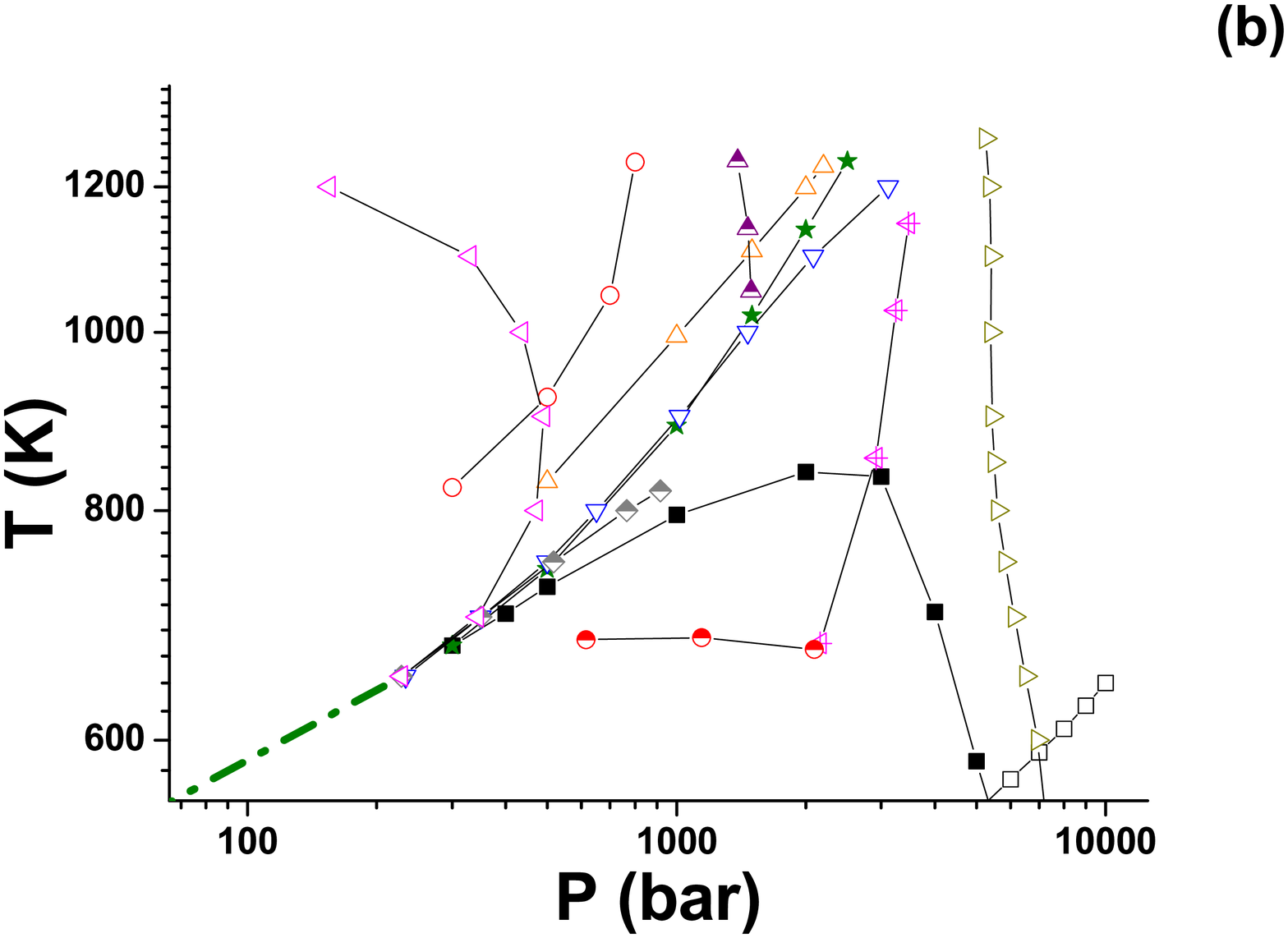}%

\includegraphics[width=6cm, height=6cm]{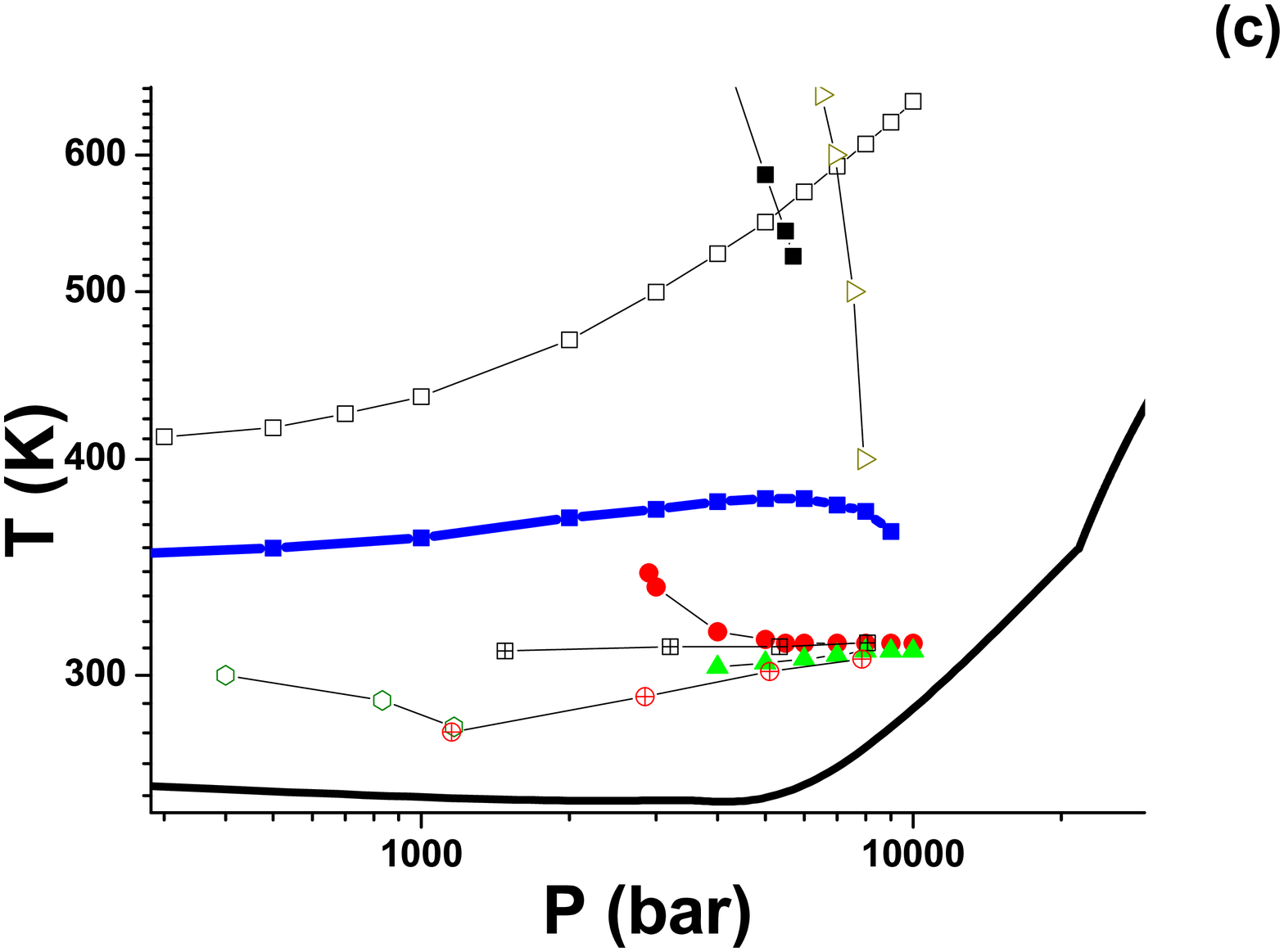}%

\caption{(a) The points of different anomalies placed on the phase
diagram of water. The labels ar the right give the quantity,
maximum of minimum and the trajectory along which the anomaly is
observed. TMD means the temperature of maximum density which
represents the density anomaly of water. Panel (b) enlarges the
region of anomalies in the vicinity of critical point. Panel (c)
enlarges the region close to the bend of the melting line. See the
text for details.}
\end{figure}

As it was discussed in the introduction, most of studies
considered or low-temperature anomalies which are located close to
the TMD line in Fig. 3 (a), or maxima in the vicinity of the
critical point (Widom lines) (Fig. 3 (b)). However, from Fig. 3
(a) one can determine a novel anomalous region of water which is
located to the higher pressures from the boiling line and mostly
below the critical temperature. From right hand side it is bounded
by the melting line. Fig. 3 (c) enlarges this region.

From Fig. 3 we can better interpret the example given above. The
maximum of $c_P$ at $T=704$ K is related to the critical point of
water, since it belongs to the Widom line of $c_P$ (black square
in the plot). However, the maximum of $c_P$ at $T=318$ K belongs
to a separate line of maxima which is located close to the bend of
the melting line and can be related to this bend. Interestingly,
only heat capacities $c_P$ and $c_V$ demonstrate maxima in the
vicinity of the bend of the melting line. The maxima of adiabatic
speed of sound $c_s$ are located at higher temperatures and can be
observed in the whole range of pressures. At moderate pressures
the maxima almost do not depend on temperature. At higher
pressures the temperature of maximum of $c_s$ passes a maximum.
However, the dependence of this maximum of the pressure is
extremely small.

Concerning the anomalies observed along isochors, many of them
fall on the boiling line and should be related to phase
transition.

We observe several anomalies of viscosity and thermal
conductivity. Minima of viscosity along isotherms are observed at
low temperatures and pressures from $P=400$ bar up to $P=1167$
bar. The location of these minima almost do not depend on
temperature ($T=300$ K at $P=400$ bar and $T=280$ K at $P=1167$
bar). Another line of viscosity minima is the line of minima on
isobars which is located in the region of Widom lines above the
critical point. Finally, there is a line of viscosity maxima along
isotherms which goes from the point $T=1275$ K and $P=5250$ bar to
the point $T=400$ K and $P=7933$ bar. This line looks to be not
connected to any special lines, like melting or boiling ones.
Interpretation of this line requires further studies.

The behavior of thermal conductivity appears to be even more
complex. In the vicinity of the critical point we observe two
lines of minima of $\lambda$ along isochors and a line of mimina
along isobars. These minima are most probably related to the
fluctuations in the vicinity of the critical point. Another line
is the line of maxima of thermal conductivity of water along
isobars which starts at $T=412$ K and $P=300$ bar and ends up at
$T=644$ K and $P=10000$ bar. Like in the case of viscosity this
line looks to be disconnected to any special lines on the phase
diagram. The interpretation of this line also requires further
studies.

\bigskip

For further investigation of the novel anomalous region of water
we performed molecular dynamics simulation. Two models of water
were used - SPC/E \cite{spce} and TIP4P \cite{tip4p}. In both
cases we simulated system of 4000 molecules in a cubic box with
periodic boundary conditions. The temperature was varied from
$T=280$ K up to $T=500$ K with step $dT=20$ K. The density varied
from $\rho=0.9$ $g/cm^3$ up to $\rho=1.66$ $g/cm^3$ with step
$0.02$ $g/cm^3$. The phase diagrams of these models of water were
calculated in several publications \cite{vega1,vega2,vega3,vega4}.
It was found that all these models overestimate the melting
pressure at a given temperature. That is why we expect that the
melting density of models should be higher with respect to the
experimental one and because of this we simulate the system up to
such extremely high densities. The time step was set to 1 fs. The
equilibration period consisted of $5 \cdot 10^6$ steps. After that
$10 \cdot 10^6$ steps were performed for production. We calculated
the equations of state and internal energy of the system at all
data points. The data were used to construct polynomial
approximations of internal energies and pressures. The
approximation for energy was used to calculate the isochoric heat
capacity $c_V$. From approximations of both internal energy and
pressure we calculated the enthalpy and used it to calculate the
isobaric heat capacity $c_P$. Our main goal was to see whether
these models demonstrate the maxima of heat capacities in the
pressure range from $P=1100$ bar up to $P=10000$ bar. All
simulations were performed using lammps simulation package
\cite{lammps}.

Fig. 4 demonstrates equations of state from NIST database and from
molecular dynamics simulations. Both TIP4P and SPC/E models are
shown. One can see that both models give rather accurate values of
pressure. Both models a bit underestimate the pressure at high
densities. SPC/E model also overestimates the pressure at the low
densities. However, the overall agrement is not bad. However, the
results for heat capacities obtained in simulations do not
describe the experimental ones at all. For example, Fig. 5 shows
isobaric and isochoric heat capacities along $P=6000$ bar isobar.
The results are given in dimensionless units, i.e. the temperature
is measured in the units of energy. One can see that TIP4P model
strongly overestimates the heat capacities. Moreover, while
experimental $c_P$ and $c_V$ demonstrate maxima, no maximum is
observed in TIP4P results. The results for SPC/E model also do not
describe the experimental data. We can conclude that common models
of water fail to reproduce the heat capacities in the high
pressure region. Therefore no computational model known up to day
is able to describe the novel region of anomalies.

\begin{figure} \label{nist-tip-spce}
\includegraphics[width=6cm, height=6cm]{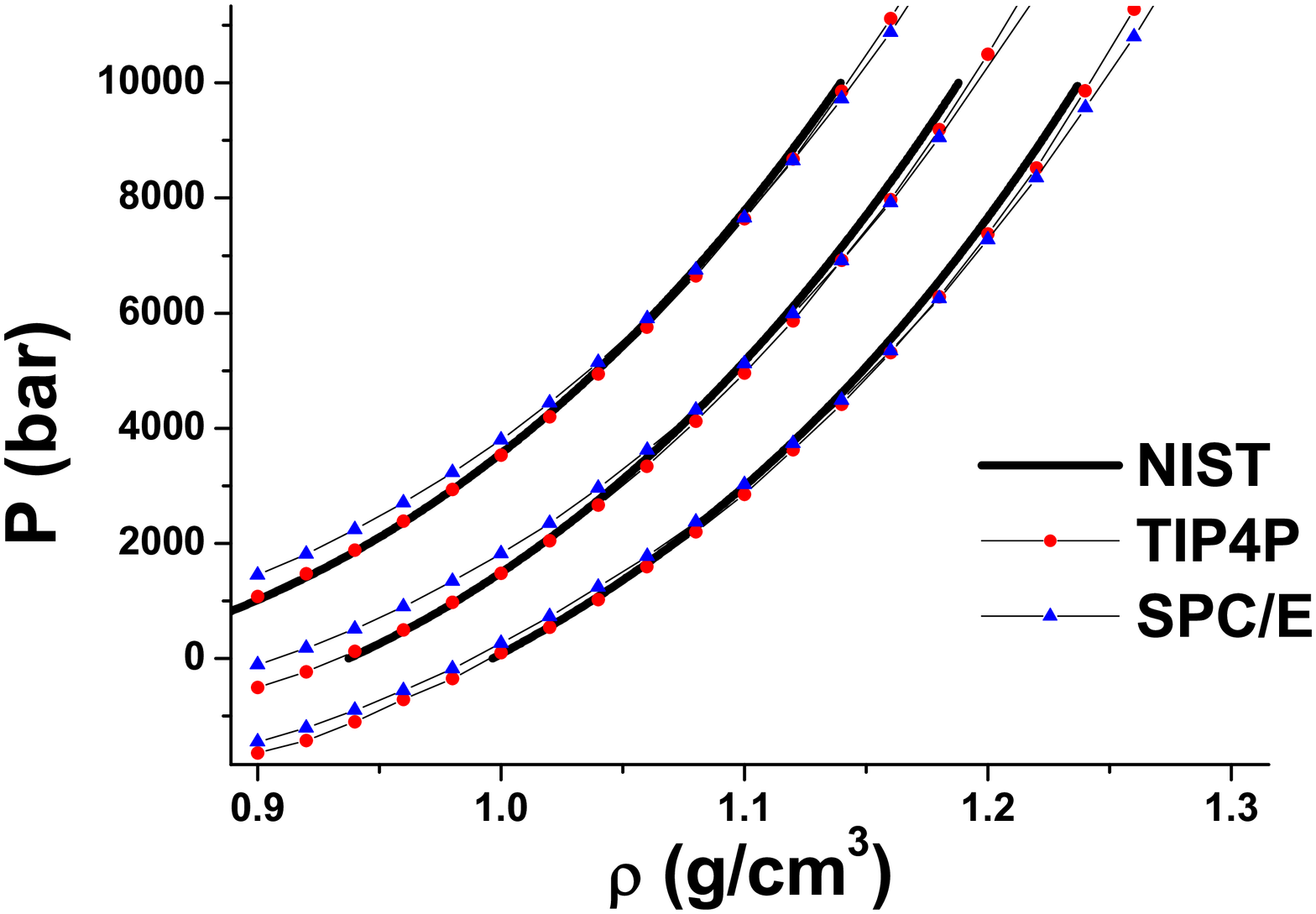}%

\caption{Comparison of equations of state from NIST database and
obtained in simulations with TIP4P and SPC/E models. The left
group of curves T=300 K, the middle one - T=400 K and the right
one - T=500 K.}
\end{figure}

\begin{figure} \label{cpcv-nist-tip}
\includegraphics[width=6cm, height=6cm]{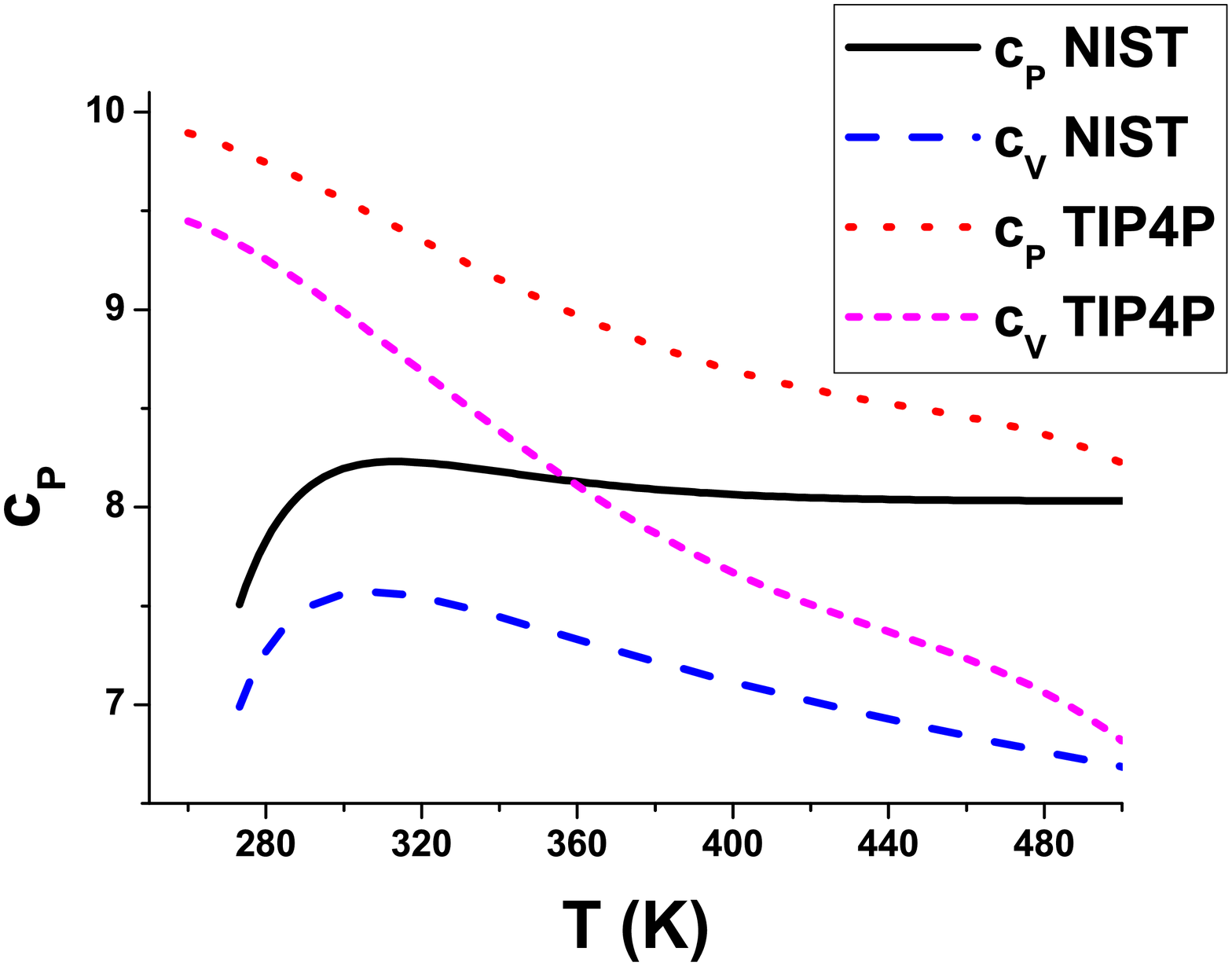}%

\caption{Comparison of heat capacities $c_P$ and $c_V$ from NIST
database and obtained in simulations with TIP4P at $P=6000$ bar
isobar.}
\end{figure}

\begin{figure} \label{regions-all}
\includegraphics[width=8cm, height=8cm]{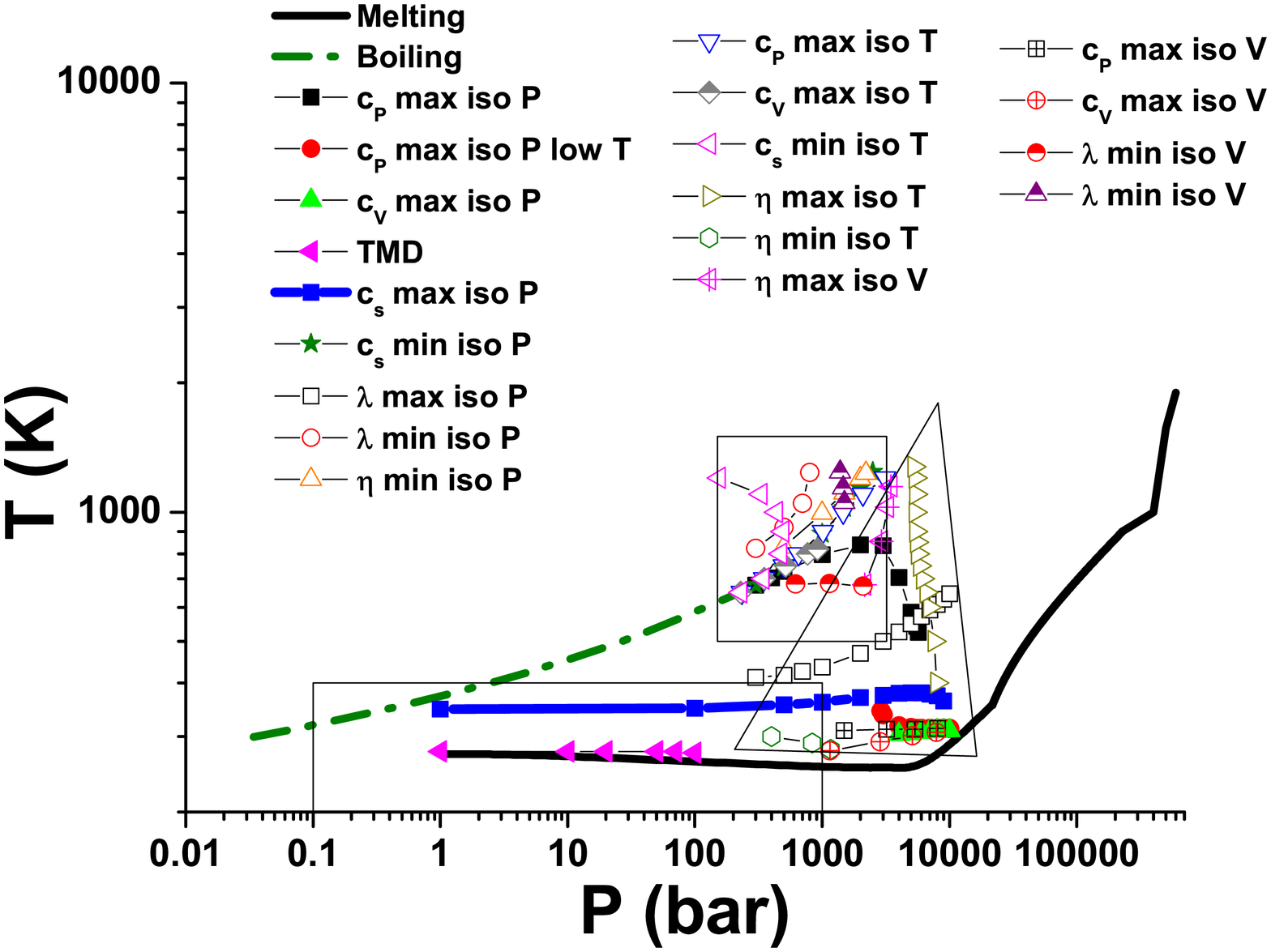}%

\caption{Phase diagram and anomalous properties of water. The
boxes show the regions where anomalous behavior can be related to
the critical points. The triangle shows the region where the
anomalous behavior cannot be explained by critical phenomena -
anomalous triangle.}
\end{figure}

\bigskip

In conclusion, we have analyzed the data on the properties of
water in a vast region of thermodynamic parameters. The results
allowed us to establish an existence of several novel anomalous
properties of water which take place in the region of high
pressure and relatively low temperature. These anomalies cannot be
related to the critical points known up to date. The region of
these anomalies can be represented as a triangle shown in Fig. 6 -
anomalous triangle. At the moment there is no any explanation of
the origin of the anomalies inside the triangle. Moreover, the
most common theoretical models fail to describe the behavior of
water in this region of pressures and temperatures. Development of
novel models of water under high pressures is required.

We thank the Russian Scientifc Center at Kurchatov Institute and
Joint Supercomputing Center of Russian Academy of Science for
computational facilities. The work was supported by the Russian
Science Foundation (Grant No 14-22-00093).

\end{document}